\documentclass[runningheads,envcountsame]{mecbic}
\providecommand{\anno}{2011}
\providecommand{\firstpage}{57}
\providecommand{\lastpage}{71}
\usepackage{graphicx}
\usepackage{amsmath}
\usepackage{amssymb}
\usepackage{enumerate}
\usepackage{color}
\usepackage{rotating}
\usepackage{multirow}
\usepackage[normalsize,tight,hang]{subfigure}

\newtheorem{theorem}{Theorem}

\newtheorem{definition}{Definition}

\newtheorem{remark}{Remark}

\newcounter{algcnt}
\Roman{algcnt}
{
  \end{trivlist}
}

\newcommand{\outdegree}{\mathtt{outdegree}}

\newcommand{\modminmin}{\mathtt{min.min}}
\newcommand{\modminmax}{\mathtt{min.max}}
\newcommand{\modmaxmin}{\mathtt{max.min}}
\newcommand{\modmaxmax}{\mathtt{max.max}}

\newcommand{\modmin}{\mathtt{min}}

\newcommand{\modmax}{\mathtt{max}}

\title{Synchronization of {P}~Systems with Simplex Channels\\ {\large (Work in Progress)}}
\def\titlerunning{Synchronization of {P}~Systems with Simplex Channels (Work in Progress)}

\author{Florentin Ipate
\institute{Department of Computer Science, University of Pite\c{s}ti,\\
T\^{a}rgul din Vale 1, Pite\c{s}ti, Romania}
\email{florentin.ipate@ifsoft.ro} \and
Radu Nicolescu
\institute{Department of Computer Science, University of Auckland,\\
Private Bag 92019, Auckland, New Zealand}
\email{r.nicolescu@auckland.ac.nz} \and
Ionu\c{t} Mihai Niculescu
\institute{Department of Computer Science, University of Pite\c{s}ti,\\
T\^{a}rgul din Vale 1, Pite\c{s}ti, Romania}
\email{ionutmihainiculescu@gmail.com} \and
Cristian \c{S}tefan
\institute{Department of Computer Science, University of Pite\c{s}ti,\\
T\^{a}rgul din Vale 1, Pite\c{s}ti, Romania}
\email{liviu.stefan@yahoo.com}
}
\def\authorrunning{F.~Ipate, R.~Nicolescu, I.M.~Niculescu and C.~\c{S}tefan}

\begin{document}

\maketitle
\pagestyle{plain}
\pagenumbering{arabic}
\setcounter{page}{57}


\begin{abstract}
We solve the Firing Squad Synchronization Problem (FSSP),
for P~systems based on digraphs with simplex channels,
where communication is restricted by the direction of structural arcs.
Previous work on FSSP for P~systems focused exclusively
on P~systems with duplex channels,
where communication between parents and children is bidirectional.
Our P~solution, the first for simplex channels,
requires cell IDs, strongly connected digraphs
and some awareness of the local topology
(such as each cell's outdegree)---we argue that these requirements are necessary.
Compared to the known solutions for cellular automata,
our solution is substantially simpler and faster.
\end{abstract}


\section{Introduction}
\label{sec:introduction}

The {\it Firing Squad Synchronization Problem} (FSSP),
originally proposed by Myhill in 1957 \cite{Moore1964},
is one of the best studied problems for cellular automata.
Essentially, the problem involves programming a network of cellular automata,
so that, after the {\it firing} order is given by the {\it general},
after some finite time,
all the cells in the system enter a designated {\it firing} state,
{\it simultaneously} and {\it for the first time}.

Versions of FSSP have been proposed and studied for variety of network structures,
from simple linear graphs to rings, trees, or general connected graphs;
see, for example,
\cite{Balzer1967,BerthiaumeBPSS2004,Goto1962,Grefenstette1983,Kobayashi1978,MooreL1968,NishitaniH1981,SchmidW-IFIPTCS2004,UmeoKNA2010,Waksman1966}.
However, most of these versions,
require {\it bidirectional} communication (i.e.~{\it duplex} channels):
only a few notable exceptions consider the more restricted {\it unidirectional} communication (i.e.~{\it simplex} channels),
starting with Kobayashi~\cite{Kobayashi1978}.
Later, Even, Litman and Winkler~\cite{Even90computingwith} 
proposed improved solutions for arbitrary {\it undirectional} {\it strongly-connected} {\it digraph},
working in $O(N^2)$ steps, where $N$ is the digraph {\it size} (number of cells).
Ostrovsky and Wilkerson~\cite{Ostrovsky95fastercomputation}
improved this further, to a solution which runs in $O(ND)$ steps,
where $D$ is the digraph {\it diameter} (typically smaller than $N$)---
this still seems to be the best solution available.

Several FSSP solutions have recently been studied in the framework of P~systems,
although with somehow different formulations,
stemming from their different computing capabilities.
P~solutions were proposed:
for {\it trees}, by Bernardini et al. \cite{BernardiniGMV2008} and Alhazov et al. \cite{AlhazovMV-WMC2008};
and for arbitrary {\it connected graphs}, by
Dinneen at al. \cite{DKN-UC2010,DKN-CMC-2011,DKN-JNC-2011}.
All these P~solutions require {\it duplex} channels and
follow the typical pattern of a {\it wave} algorithm~\cite{Tel2000},
using three phases:
\begin{enumerate}
\item a {\it first broadcast}---which follows all shortest paths from the {\it general} and builds a {\it virtual BFS tree} (or {\it dag});

\item a {\it convergecast}---which helps determine the {\it general's eccentricity};

\item a {\it second broadcast}---which carries the actual firing command
(with a countdown counter).
\end{enumerate}

The best P~solutions need $e_g+k$ steps for each of the three phases,
for a total of $3e_g+k$ steps, where~$e_g$ is the general's eccentricity
(height for trees, if the general is at the root).

Obviously, while the two broadcasts, of phases (1) and (3),
would also work with simplex channels,
duplex channels are essential for the convergecast of phase (2),
where children need to talk back to their parents.
At first sight, the convergecast seems impossible for simplex channels.
However, children can still talk back to their parents,
if the digraph is {\it strongly-connected},
albeit on a typically longer {\it path}.
Moreover, if messages cannot be confused,
all children can send messages to their parents,
in parallel (overlapping in time without problems),
achieving this way a {\it virtual convergecast}.

Based on ideas from Ostrovsky and Wilkerson~\cite{Ostrovsky95fastercomputation},
we propose a {\it first} FSSP solution for P~systems
with {\it simplex} channels, based on arbitrary {\it strongly-connected digraphs}.
Our solution runs in $O(e_g D)$ steps, specifically:
(1) $e_g$ steps for the first broadcast;
(2) $O(e_g D)$ steps for the virtual convergecast (maximally parallelized);
(3) $e_g$ steps for the second broadcast.
Thus, in terms of execution time,
our P~solution compares favourably with the best known solution for cellular automata.
Taking into account the different problem constraints
and different computing capabilities of P~systems vs. cellular automata,
we were expecting a simpler and faster solution,
but not necessarily such a substantial speed improvement.
However, for the reasons mentioned,
any performance comparison must be viewed with a grain of salt.

The actual design seems challenging and requires careful selection of the most adequate ingredients, some of which are available in cellular automata, but not typically available in P~systems.
Specifically, we argue that the P~solution requires: (1) reified cell IDs;
(2) reified local network information, such as the number of outgoing arcs (this information {\it is} available in cellular automata);
and (3) high-level generic rules, if we want to have a fixed rule set,
independent of the actual network size.


\section{Preliminaries}
\label{sec:preliminaries}

We assume that the reader is familiar with the basic terminology and notations,
such as relations, graphs, nodes (vertices), edges, directed graphs (digraphs),
directed acyclic graphs (dags), arcs, alphabets, strings and multisets.

A {\it P~system} is a parallel and distributed computational model,
inspired by the structure and interactions of cell membranes.
This model was introduced by P\u{a}un in 1998--2000 \cite{Paun2000}.
An in-depth overview of this model can be found in P\u{a}un et al. \cite{Paun2010}.

In this paper, we consider an {\it ad-hoc} definition of P~systems,
based on our definition of {\it simple P~module}~\cite{DKN-EPTCS2010},
which extends earlier versions of tissue and neural P~systems~\cite{MartinVidePPR2003,Paun2002}.
However, here we intentionally {\it restrict} rule transfer mode to broadcast to all children, ${\downarrow_\forall}$.

\begin{definition}
\label{def-P-system}
A {\it P~system} of order $n$ with {\it simplex channels}
is a system $\Pi = (O, K, \delta)$, where:
\begin{enumerate}
  \item $O$ is a finite alphabet of {\it elementary symbols}; strings over $O$ are interpreted as multisets;
  \item $K = \{\sigma_1, \sigma_2, \dots, \sigma_n\}$ is a finite set of {\it cells};
        where each cell is a system $\sigma_i = (Q_i, R_i)$,
        with $Q_i$ a finite set of {\it states}
        and $R_i$ a finite set of rewriting rules over $O$, further detailed below.
  \item $\delta$ is an {\it irreflexive} binary relation on $K$,
        which represents a set of structural arcs between cells,
        with {\it unidirectional} communication capabilities,
        strictly from parents to children.
  \item Each $R_i$ is a finite {\it linearly ordered} set of multiset rewriting {\it rules}
        with {\it promoters}, of the form:
        $S~x \rightarrow_\alpha S'~x'~(y)_{\downarrow_\forall} \dots \mid_z$,
        where $S, S' \in Q_i$, $x, x', y \in O^*$, $z \in O^*$ is the promoter,
        $\alpha \in \{\modmin, \modmax\}$ is a {\it rewriting} operator and
        ${\downarrow_\forall}$ is a {\it transfer} operator,
        here restricted to send $y$ messages from a parent to all its children.
\end{enumerate}
\end{definition}

As usually, each cell, $\sigma_i \in K$, starts from
an {\it initial configuration} $(S_{i0}, w_{i0})$,
where $S_{i0} \in Q_i$ is its {\it initial state}
and $w_{i0} \in {O}^*$ is its {\it initial content}.
A cell {\it evolves} by applying one or more rules,
which can change its {\it current configuration},
i.e.~its current state and current content,
and send symbols to its children.

The application of a rule transforms the {\it current state} $S$
to the {\it target state} $S'$,
rewrites multiset $x$ as $x'$ and sends multiset $y$ by replication to all its children.
Note that, multisets $x'$ and $y$ will not be visible to further rules in this same step, but they will become visible after no more rules are applicable,
i.e.~they will be available since next step only.
{\it Promoters} are symbols which enable rules, but are not consumed by the rules' application.

When an applicable rule is applied,
its rewriting operator $\alpha$ indicates
how many times it is actually applied:
{\it once}, if $\alpha = \modmin$;
or {\it as many times as possible}, if $\alpha = \modmax$.

As used here, rules have {\it priorities} and
are applied in {\it weak priority} order~\cite{Paun2006},
with special attention to target state compatibility:
(1) higher priority applicable rules are applied before
lower priority applicable rules, and
(2) a lower priority applicable rule is applied
only if it indicates the same target state as the previously assigned rules (if any).

All cells evolve {\it synchronously} in one global {\it step}.
An {\it evolution} of a P~system is a sequence of steps,
where each cell starts from its initial configuration.
An execution {\it halts} if no cell can evolve.

\subsection{Further P~systems extensions}
\label{sec:extensions}

We let each cell, $\sigma_i$, start with its own unique {\it cell ID} symbol, $\iota_i$. We thus {\it reify} the conceptual cell index,~$i$, into an internal
symbol, which is accessible to the rules, exclusively as
an {\it immutable promoter}~\cite{NW-UC-2011}.

We enhance our vocabulary by recursive composition
of {\it elementary symbols} from $O$ into a simple form
of {\it complex symbols}~\cite{NW-UC-2011}.
Such complex symbols can be viewed as complex molecules,
consisting of elementary atoms or other molecules.

Further, complex symbols let us process our multisets
with high-level {\it generic rules},
using {\it free variable} matching.
To explain these additional ingredients, consider this hypothetical rule
(which uses an additional transfer mode, targeted to a specific child,
not considered in the definition used here):


$$S ~a ~n_j \rightarrow_{\modminmin} S' ~b ~(c_i)_{\downarrow_j} ~| \iota_i.$$

This is a {\it generic} rule, which uses an {\it extended rewriting mode},
with complex symbols, $c_i$ and $n_j$, where $i$ and $j$ are free variables.
In fact, $c_i$ and $n_j$ are just shorthands for {\it tuples} $(c, i)$ and $(n, j)$,
or, equivalently, for {\it compound terms} $c(i)$ and $n(j)$.
If needed, we can build more complex symbols by recursive composition;
e.g., we could have complex symbols such as $d(e,i,f(j))$.
Generally, a free variable could match anything, including another complex symbol.
However, in this rule, $i$ and $j$ are constrained to match cell ID indices only:
\begin{enumerate}
\item $i$---because it also appears as the cell ID of the current cell, $\iota_i$;
\item $j$---because it also indicates the target of the transfer mode, $\downarrow_j$.
\end{enumerate}

A generic rule is identified by using an {\it extended} version of the ``classical'' rewriting mode, in fact, by a combined {\it instantiation} and {\it rewriting} mode.
Our sample rule uses the extended mode $\modminmin$,
where the two $\modmin$ operators have distinct semantics:
the first $\modmin$ operator is new and describes the generic instantiation;
the second $\modmin$ is the classical operator, which describes the rule application.
Briefly:
\begin{enumerate}
\item according to the first $\modmin$, this rule is {\it instantiated} {\it once},
for one of the existing $n_j$ symbols (if any), while promoter, $\iota_i$,
constrains $i$ to the cell ID index of the current cell, $\sigma_i$;

\item according to the second $\modmin$,
the instantiated rule is {\it applicable} {\it once},
i.e.~if applied, it consumes one $a$ and one $n_j$,
produces one $b$ and sends one $c_i$ to child $\sigma_j$
(if this exists).
\end{enumerate}

As a further example, consider the scenario in which the current cell, $\sigma_1$,
contains the multiset $n_2 n_3 n_3$.
Here, our sample generic rule instantiates (randomly)
{\it one} of the following two lower-level rules,
which is then applied in the classical way (in the $\modmin$ rewriting mode):

$$S ~a ~n_2 \rightarrow_{\modmin} S' ~b ~(c_1)_{\downarrow_2}.$$
$$S ~a ~n_3 \rightarrow_{\modmin} S' ~b ~(c_1)_{\downarrow_3}.$$

We consider four basic combinations
of the instantiation and rewriting modes,
$\modminmin$, $\modminmax$, $\modmaxmin$, $\modmaxmax$;
their semantics is:

\begin{itemize}
\item $\modminmin$ indicates that
the generic rule is (randomly) instantiated {\it once}, if possible,
and the instantiated rule is applied {\it once}, if possible.

\item $\modminmax$ indicates that
the generic rule is (randomly) instantiated {\it once}, if possible,
and the instantiated rule is applied as {\it many} times as possible.

\item $\modmaxmin$ indicates that
the generic rule is instantiated as {\it many} times as possible,
without superfluous instances
(i.e.~without duplicates or instances which are not applicable),
and each one of the instantiated rules is applied {\it once}, if possible.

\item $\modmaxmax$ indicates that
the generic rule is instantiated as {\it many} times as possible,
without superfluous instances
(i.e.~without duplicates or instances which are not applicable),
and each one of the instantiated rules is applied as {\it many} times as possible.
\end{itemize}

All instantiations
are ephemeral, created when rules
are tested for applicability and disappearing at the end of the step.


\section{FSSP problem for P~systems with simplex channels}
\label{sec:problem}

We are required to find:
\begin{enumerate}
\item an alphabet $O$;
\item a {\it cell prototype} $\sigma = (Q, R)$, where
	\begin{enumerate}
     \item R is a set of rules over $O$;
	\item $Q$ contains two distinguished states:
           \begin{itemize}
           \item $S_0$: a {\it quiescent} state,
           i.e. if $\sigma$ is in state $S_0$ and {\it empty}, then there are no applicable rules;
           \item $S_f$: a {\it final} state,
           i.e.~if $\sigma$ is in state $S_f$, then there are no applicable rules.
           \end{itemize}
      \end{enumerate}
\end{enumerate}

\noindent such that, given:
\begin{enumerate}
\item {\it any} finite set of $\sigma$ copies,
$K = \{\sigma_1, \sigma_2, \dots, \sigma_n\}$, $\sigma_i = \sigma$;
\item connected via {\it any} strongly-connected digraph $\delta$;
\end{enumerate}

\noindent the P~system $\Pi = (O, K, \delta)$ with {\it simplex channels}
will evolve according to the following specification:
\begin{enumerate}
\item all cells start from quiescent state $S_0$: $S_{i0} = S_0$;
\item except a distinguished cell $\sigma_g$, called the {\it general},
all cells start with a restricted initial content, containing,
the reified cell ID and a reified count of the cell's outdegree:
$w_{i0} \subseteq \{ \iota_i , c^{\outdegree(\sigma_i)} \}, \forall i \neq g$;
\item the evolution terminates and, during its last step:
      all cells enter state $s_f$ {\it simultaneously} and {\it for the first time}.
\end{enumerate}

\begin{remark}
Our formulation has different constraints than the original problem
for cellular automata.
In the cellular automata formulation,
there is a given {\it fixed bound} on the number
of input and output connections of each cell (bounded indegree and outdegree).
In our formulation, there is {\it no bound} on the number of
input and output connections that a cell may have.
However, this is compensated by the fact that in our formulation
there are {\it no size} bounds on messages or cells' internal memory.
These {\it trade-offs}, as well as different computing capabilities,
suggests that performance comparisons must be viewed with a grain of salt.
\end{remark}

\begin{remark}
We argue that both the reified cell ID and the reified children count,
or equivalent information, are necessary, definitely for our approach,
and, likely, for any other approach.
Note that children counts are implicit in the cellular automata version,
where unconnected channels (out of the fixed sized pool) can be detected.
\end{remark}

\begin{remark}
There is no constraint on the cells' final contents.
However, if needed, any left-over garbage could be collected in one extra step.
\end{remark}

\begin{remark}
Practically, we are only required to design the rule set, $R$,
because this implies the alphabet,~$O$, and the state set, $Q$.
\end{remark}

\begin{remark}
Note the rule set, $R$, must be fixed and applicable
to any structural digraph.
This is a strong requirement:
we require a rule set which is
independent of the size and structure of the actual system.
\end{remark}


\section{FSSP solution for P~systems with simplex channels}
\label{sec:digraphs}


Our solution runs in three phases (conceptually similar to the duplex case):

\begin{enumerate}
\item
First phase: a {\it first broadcast} from the general.
This phase builds the {\it virtual-dag} (the virtual BFS dag)
and, for each cell:
(a) records its {\it virtual-dag-parent(s)};
and (b) successively computes its {\it depth} attribute,
which represents this cell's depth level in the virtual-dag 
(the same as this cell's digraph distance from the general).
A first phase broadcast message is a complex symbol, $x_{k,i}$,
where $i$ is the sender's ID and $k$ is the next {\it depth} level
($\sigma_i$'s own {\it depth} plus one).

\item
Second phase: a {\it virtual convergecast} from the virtual-dag leaves.
For each cell, this phase successively computes the {\it max-depth} attribute,
which represents the maximum depth over all descendant cells in the virtual-dag.
In the end, the general's {\it max-depth} is its eccentricity.

This virtual convergecast simulates impossible direct virtual-dag-child to virtual-dag-parent messages, by broadcasting them over the digraph
(using ad-hoc BFS dags).
A convergecast message is a complex symbol, $a_{j,i,k}$,
where $i$ is the sender's ID, $j$ is its virtual-dag-parent ID 
and $k$ is $\sigma_i$'s own {\it max-depth}.
In addition to virtual-dag leaves, a pseudo-convergecast message is also
sent by a digraph cell to its digraph parents with 
larger {\it depth} attributes (if any). 
Because each message is uniquely identified by both it sender
and its destination,
any number of such convergecasts can run in parallel, without creating confusions.

Note that a cell {\it needs} to know its digraph outdegree---to 
detect when it receives its last outstanding convergecast message 
and to start its own convergecast.
However, a cell does {\it not} know the identities of its digraph children,
or how long a message from any one of them will take to reach it.

\item
Third and last phase: a {\it second broadcast} from the general, with a countdown to firing.
A last phase broadcast message is a complex symbol, $f_k$,
where $k$ is the next countdown counter
($\sigma_i$'s own countdown minus one).

\end{enumerate}

As possible extensions, not discussed here, 
we can extract from our solution a more general subprogram,
to send any message from any cell to any other cell, 
which can run in parallel, without creating confusions.
Also, we can consolidate the routing information, to speedup
future messages with the same destination;
however, this feature is not required here
(each convergecast is performed exactly once).

\begin{figure}[h]
\centerline{\includegraphics[scale=1.5]{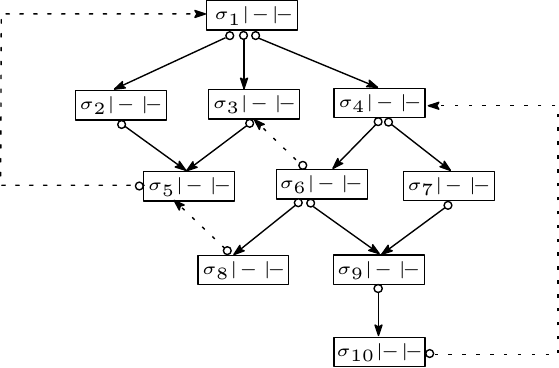}}
\caption{
$\Pi$: Initial configuration of a sample P~system with simplex channels,
based on a strongly connected digraph.
{\small Normal arrows are arcs in the virtual dag,
created by a BFS broadcast, started from the general, $\sigma_1$.
The remaining digraph arcs are dotted arrows.
Each cell knows its digraph outdegree,
indicated by small circles at outgoing arrows' tails.
Each cell blob shows three attributes, in order:
(1) its cell ID;
(2) its {\it depth} attribute (computed by the first broadcast);
and (3) its {\it max-depth} attribute (computed by the virtual convergecast).
At this stage, the {\it depth} and {\it max-depth} attributes are still indeterminate
and all small circles are white, indicating still outstanding convergast messages.
}}
\label{fig:sample-simplex}
\end{figure}

\begin{figure}[h]
\centerline{\includegraphics[scale=1.5]{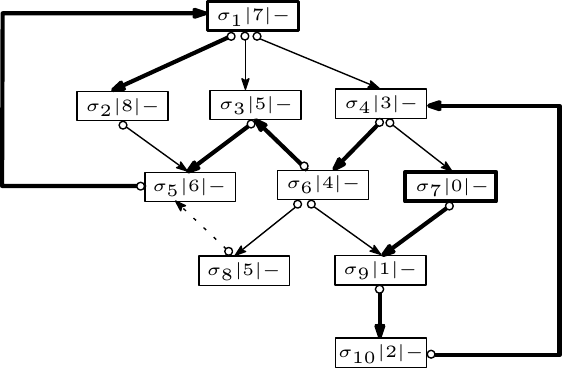}}
\caption{
$\Pi$: route from $\sigma_7$ to $\sigma_2$,
indicated via thick arrows.
}
\label{fig:sample-route}
\end{figure}

\begin{figure}[h]
\centerline{\includegraphics[scale=1.5]{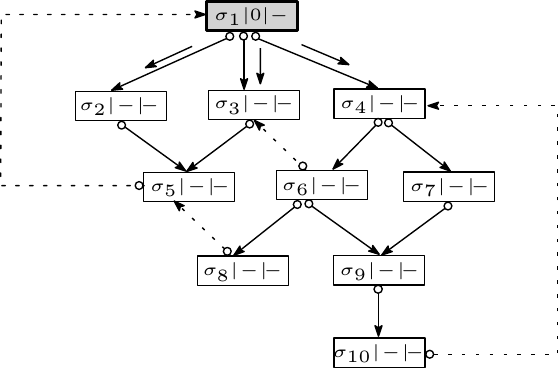}}
\caption{
$\Pi$, step 1:
The general, $\sigma_{1}$, starts the first phase,
by broadcasting the complex symbol $a_{1,1}$.
}
\label{fig:step-1}
\end{figure}

\begin{figure}[h]
\centerline{\includegraphics[scale=1.5]{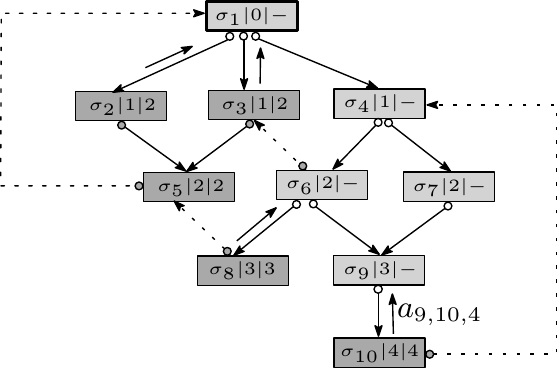}}
\caption{$\Pi$, step 9:
Cell $\sigma_{10}$ has just learned that it is a virtual-dag leaf, 
by receiving cell's $\sigma_4$'s pseudo-convergecast message, the complex symbol $a_{10,4,-}$.
Cell $\sigma_{10}$ starts now its virtual convergecast, 
by broadcasting the complex symbol $a_{9,10,4}$,
towards its virtual-dag-parent, $\sigma_9$.
{\small At this stage, each cell knows its virtual-dag-parent(s) 
and its own {\it depth} attribute.
All other cells have already initiated their virtual convergecasts.
All leaves, including $\sigma_{10}$, and some other cells already know their {\it max-depth} attribute:
exactly, if they have received all their outstanding convergecast messages,
or a lower bound, otherwise.
Message $a_{9,10,4}$ will take three more steps:
via paths $\sigma_{10}.\sigma_4.\sigma_6.\sigma_9$
and path $\sigma_{10}.\sigma_4.\sigma_7.\sigma_9$.
Three other, earlier stared, virtual convergecasts run in parallel:
$\sigma_2$ to $\sigma_1$, $\sigma_3$ to $\sigma_1$, $\sigma_8$ to $\sigma_6$.
Smaller arrows near structural arcs indicate virtual convergecasts.
}}
\label{fig:step-9}
\end{figure}

\begin{figure}[h]
\centerline{\includegraphics[scale=1.5]{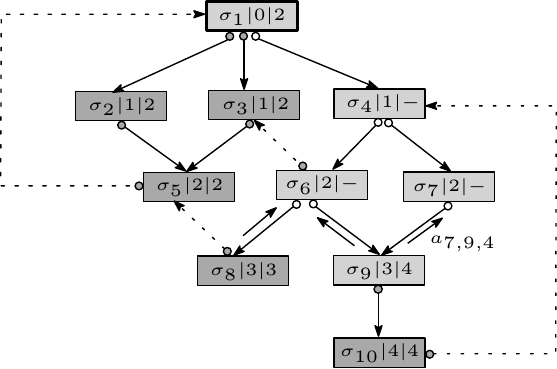}}
\caption{
$\Pi$, step 12:
After receiving its single expected convergecast message,
cell $\sigma_{9}$ starts its own virtual convergecasts towards its parents, $\sigma_6$ and $\sigma_7$,
by broadcasting complex symbols, $a_{6,9,4}$ and $a_{7,9,4}$, respectively.
{\small Each of these messages will take three more steps,
via path $\sigma_{9}.\sigma_{10}.\sigma_4.\sigma_6$
and via path $\sigma_{9}.\sigma_{10}.\sigma_4.\sigma_7$, respectively.
These two convergecasts run in parallel with another virtual convergecast: $\sigma_8$ to $\sigma_6$.
}}
\label{fig:step-12}
\end{figure}

\begin{figure}[h]
\centerline{\includegraphics[scale=1.5]{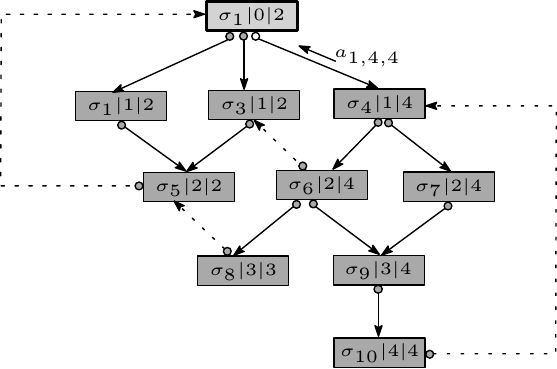}}
\caption{
$\Pi$, step 18:
The convergecast phase is almost completed.
Cell $\sigma_{4}$ starts its convergecast towards its parent, $\sigma_1$,
by broadcasting the complex symbol $a_{1,4,4}$.
{\small This convergecast will take four more steps,
via paths $\sigma_4.\sigma_6.\sigma_3.\sigma_5.\sigma_1$
and path $\sigma_4.\sigma_6.\sigma_8.\sigma_5.\sigma_1$.
This is cell $\sigma_1$'s last oustanding convergecast.
}}
\label{fig:step-18}
\end{figure}

\begin{figure}[h]
\centerline{\includegraphics[scale=1.5]{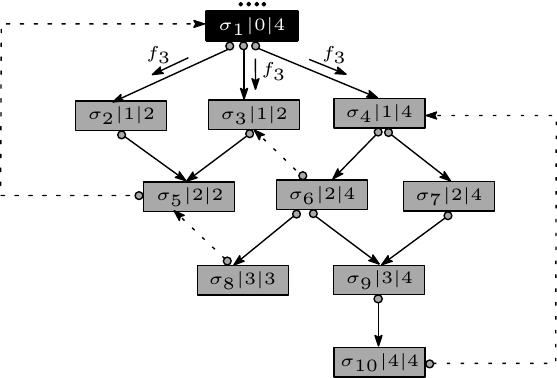}}
\caption{
$\Pi$, step 22:
General $\sigma_1$ starts the last phase, by broadcasting
the complex symbol $f_3$, carrying the countdown to firing.
{\small Small dots above a cell indicate the cell's countdown counter,
which is also broadcasted to all its digraph children.
}}
\label{fig:step-22}
\end{figure}

\begin{figure}[h]
\centerline{\includegraphics[scale=1.5]{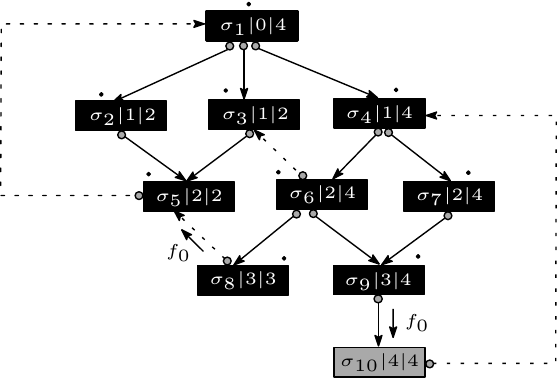}}
\caption{
$\Pi$, step 25:
The last phase is almost complete.
Cell $\sigma_9$ forwards the complex symbol $f_0$ to $\sigma_{10}$.
{\small This the last step before firing (not illustrated here).
}}
\label{fig:step-25}
\end{figure}

Figures~\ref{fig:sample-simplex}--\ref{fig:step-25}
show bird's eye views of the evolution of $\Pi$:
a sample P~system, with simplex channels,
based on a strongly connected digraph.

\medskip


\subsection{Rule set $R$}
\label{sec:ruleset}


\begin{tabular}[t]{ll}
  \begin{minipage}[t]{2.7in}
  \begin{enumerate}
  \setcounter{enumi}{-1}

  \item Rules in state $S_0$:
    \begin{enumerate}[1.]
    \item $S_0~ a \rightarrow_\modminmin \\ ~~~~~~~~~~ S_1~ g~ n_0~ m_0~ (x_{1,i})_{\downarrow_\forall}~ | \iota_i$

    \item $S_0~ x_{k,j} \rightarrow_\modmaxmin \\ ~~~~~~~~~~~~~~ S_1~ l_k~ p_j~ (x_{k+1,i})_{\downarrow_\forall}~ | \iota_i$
    \item $S_0~ x_{k,j} \rightarrow_\modmaxmax S_1~p_j~ | \iota_i$
    \end{enumerate}

  \vspace{2mm}
  \item Rules in state $S_1$:
    \begin{enumerate}[1.]
    \item $S_0~ l_k \rightarrow_\modmaxmin S_1~ n_k~ m_k$
    \item $S_0~ l_k \rightarrow_\modmaxmax S_1$

    \item $S_1~ x_{k,j} \rightarrow_\modmaxmin S_1~ y_{j,0}$
    \item $S_1~ x_{k,j} \rightarrow_\modmaxmax S_1$

    \item $S_1~ s_k \rightarrow_\modmaxmin Max~ w~ m_k$

    \item $S_1~ y_{j,k} \rightarrow_\modmaxmin \\ ~~~~~~~~~~~~~ S_1~ v_{j,i}~ (a_{j,i,k})_{\downarrow_\forall}~ | \iota_i$
    \item $S_1~ a_{j,k,l} \rightarrow_\modmaxmax S_1~ | v_{j,k}$
    \item $S_1~ a_{i,j,k} \rightarrow_\modmaxmin S_1~ v_{i,j}~ s_k~ | \iota_i$
    \item $S_1~ a_{j,k,l} \rightarrow_\modmaxmin \\ ~~~~~~~~~~~~~~~ S_1~ v_{j,k}~ (a_{j,k,l})_{\downarrow_\forall}$
    \end{enumerate}

  \vspace{2mm}
  \item Rules in state $S_2$:
    \begin{enumerate}[1.]
    \item $S_2~ t \rightarrow_\modmin S_5~ (t)_{\downarrow_\forall}$

    \item $S_2~ s_k \rightarrow_\modmaxmin Max~ w~ m_k$

    \item $S_2~ y_{j,k} \rightarrow_\modmaxmin \\ ~~~~~~~~~~~~~ S_2~ v_{j,i}~ (a_{j,i,k})_{\downarrow_\forall}~ | \iota_i$
    \item $S_2~ a_{j,k,l} \rightarrow_\modmaxmax S_2~ | v_{j,k}$
    \item $S_2~ a_{i,j,k} \rightarrow_\modmaxmin S_2~ v_{i,j}~ s_k~ | \iota_i$
    \item $S_2~ a_{j,k,l} \rightarrow_\modmaxmin \\ ~~~~~~~~~~~~~~~ S_2~ v_{j,k}~ (a_{j,k,l})_{\downarrow_\forall}$
    \end{enumerate}

  \vspace{2mm}
  \item Rules in state $S_3$:
    \begin{enumerate}[1.]
    \item $S_3~ c c \rightarrow_\modmin S_1~ c e$
    \item $S_3~ c \rightarrow_\modmin S_4~ e b$
  \end{enumerate}

  \end{enumerate}
  \end{minipage}

  \begin{minipage}[t]{2.2in}
  \begin{enumerate}
  \setcounter{enumi}{2}

  \item Rules in state $S_4$:
    \begin{enumerate}[1.]

    \item $S_4~ \rightarrow_\modmaxmin \\ ~~~~~~~~~ S_2~ t~ f_k~ (t)_{\downarrow_\forall}~ | g~b~m_k$
    \item $S_4~ \rightarrow_\modmaxmin S_2~ y_{j,k}~ | p_j~ m_k$
    \end{enumerate}

  \vspace{2mm}
  \item Rules in state $Max$ (needs refinement):
    \begin{enumerate}[1.]
    \item $Max~ m_k \rightarrow_\modmax S_3~ | m_{k+l}$
  \end{enumerate}

  \vspace{2mm}
  \item Rules in state $S_5$:
    \begin{enumerate}[1.]
    \item $S_5 ~ v_{k,l} \rightarrow_\modmaxmax S_f~ | f_0$
    \item $S_5 ~ p_k \rightarrow_\modmaxmax S_f~ | f_0$
    \item $S_5 ~ n_k \rightarrow_\modmaxmax S_f~ | f_0$
    \item $S_5 ~ m_k \rightarrow_\modmaxmax S_f~ | f_0$
    \item $S_5 ~ e \rightarrow_\modminmax S_f~ | f_0$
    \item $S_5 ~ b \rightarrow_\modminmax S_f~ | f_0$
    \item $S_5 ~ \rightarrow_\modminmin S_f~ | f_0$

    \item $S_5 ~ v_{k,l} \rightarrow_\modmaxmax S_5 ~ | f_k$
    \item $S_5 ~ p_k \rightarrow_\modmaxmax S_5 ~ | f_k$
    \item $S_5 ~ n_k \rightarrow_\modmaxmax S_5 ~ | f_k$
    \item $S_5 ~ m_k \rightarrow_\modmaxmax S_5 ~ | f_k$
    \item $S_5 ~ e \rightarrow_\modminmax S_5 ~ | f_k$
    \item $S_5 ~ b \rightarrow_\modminmax S_5 ~ | f_k$

    \item $S_5 ~ f_k \rightarrow_\modmaxmin \\ ~~~~~~~~~~~ S_5 ~ f_{k-1}~ (f_{k-1})_{\downarrow_\forall}$
    \item $S_5 ~ f_k \rightarrow_\modmaxmax S_5 ~$

    \item $S_5 ~ t \rightarrow_\modmax S_5 ~$
  \end{enumerate}

  \end{enumerate}
  \end{minipage}
\end{tabular}


\subsection{Alphabet $O$, elementary and complex symbols}
\label{sec:symbols}


\begin{itemize}
\item $\iota_i$: reified cell ID;

\item $a$: starts the process from the cell which next assume the general role;

\item $g$: marks the general;

\item $x_{j,k}$: complex symbol broadcasted in the first phase;

\item $n_k$: indicates the {\it depth} attribute, $k$;

\item $m_k$: indicates the {\it max-depth} attribute, $k$;

\item $l_k$: auxiliary symbol used to compute {\it depth} attribute, $n_k$;

\item $p_k$: pointer to a parent cell;

\item $y_{i,k}$: initiates the sending of message $k$ to $\sigma_i$;

\item $a_{i,j,k}$: complex symbol broadcasted in the convergecast phase, sent from $\sigma_j$ to $\sigma_i$ and carrying payload $k$, 
representing $\sigma_j$'s {\it max-depth} 
(if known, otherwise it is a pseudo-convergcast sent by a cell with lower {\it depth});

\item $v_{i,j}$: records the passage of a message from $\sigma_j$ to $\sigma_i$;

\item $s_k$: auxiliary symbol used to compute {\it max-depth} attribute, $m_k$, via maximum;

\item $c$: used to count the children which have not yet sent their convergecast messages (initially cell's outdegree);

\item $e$: used to count the children which have already sent their convergecast messages (initially zero);

\item $b$ : marks a cell which has performed its convergecast;

\item $t$: auxiliary symbol used in the countdown to firing;

\item $f_k$: complex symbol broadcasted in the last phase, carrying the countdown to firing;
\end{itemize}


\subsection{Brief description}
\label{sec:description}


\begin{enumerate}
\item State $S_0$: initiates the first broadcast and computes {\it depth} attributes.

\item State $S_1$: continues the first broadcast and builds the virtual-dag.

\item State $S_2$: performs the actual convergecast.

\item State $S_3$: decides if a non-general cell is ready to start its convergecast, i.e.~if it has received its all outstanding convergecast messages, from all its digraph children.

\item State $S_4$: decides if the general is ready to start the second broadcast (countdown to firing), i.e.~if it has received convergecast messages from all its children.

\item State $Max$: prepares the convergecast, by determining the {\it max-depth} attribute. 

\item State $S_5$: last broadcast, counts down to firing and erases unnecessary symbols.
\end{enumerate}


\section{Assessment}
\label{sec:assessment}


\begin{theorem}
\label{thm-sync-time-digraphs}
The synchronization time of the FSSP solution for digraph-based P~systems
with simplex channels is
$e_g + e_g D + e_g$, i.e.~bounded by $O(e_g D)$.

\end{theorem}

\begin{theorem}
The digraph must be strongly connected, otherwise, there is no solution.

\end{theorem}

\begin{theorem}
Cells must know the number of their outgoing arcs, otherwise, there is no solution.

\end{theorem}

\begin{theorem}
Cell IDs must be reified, otherwise, there is no solution.

\end{theorem}


\section{Experimental results}
\label{sec:experimental}

\begin{figure}[ht]
\centering
\begin{tabular}{cc}

\begin{minipage}{1.5in}
\subfigure[Sample network, $N=6$.]{
\includegraphics[scale=0.5]{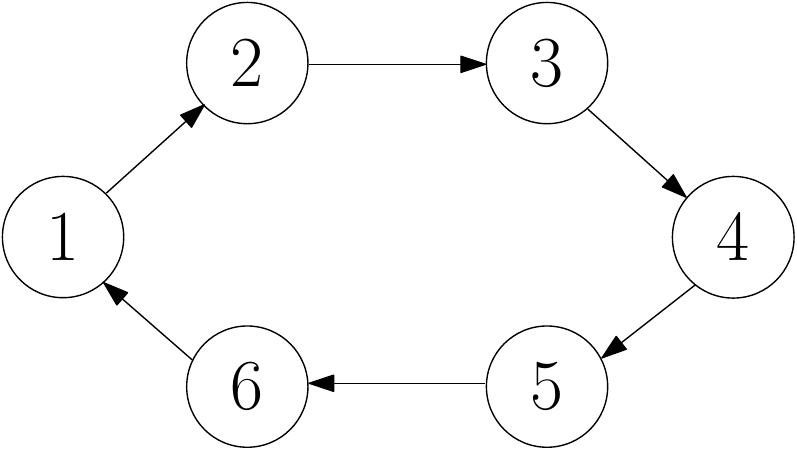}
\label{fig:fring1}
}
\end{minipage}
&
\begin{minipage}{1.5in}
\subtable[Results.]{
\begin{tabular}{|c|c|c|c|}
\hline
$N$ & $e_g$ & $D$ & Steps \\ \hline
2  &  1 &  1 &  18 \\ \hline
3  &  2 &  2 &  29 \\ \hline
4  &  3 &  3 &  42 \\ \hline
5  &  4 &  4 &  57 \\ \hline
6  &  5 &  5 &  74 \\ \hline
7  &  6 &  6 &  93 \\ \hline
8  &  7 &  7 & 114 \\ \hline
9  &  8 &  8 & 137 \\ \hline
10 &  9 &  9 & 162 \\ \hline
11 & 10 & 10 & 189 \\ \hline
12 & 11 & 11 & 218 \\ \hline
13 & 12 & 12 & 249 \\ \hline
14 & 13 & 13 & 282 \\ \hline
15 & 14 & 14 & 317 \\ \hline
\end{tabular}
\label{tbl:tring1}
}
\end{minipage}

\end{tabular}
\caption{Ring networks.}
\label{fig:ring1}
\end{figure}
\begin{figure}[ht]
\centering
\begin{tabular}{cc}

\begin{minipage}{2in}
\subfigure[Sample network, $N=10$.]{
\includegraphics[scale=0.4]{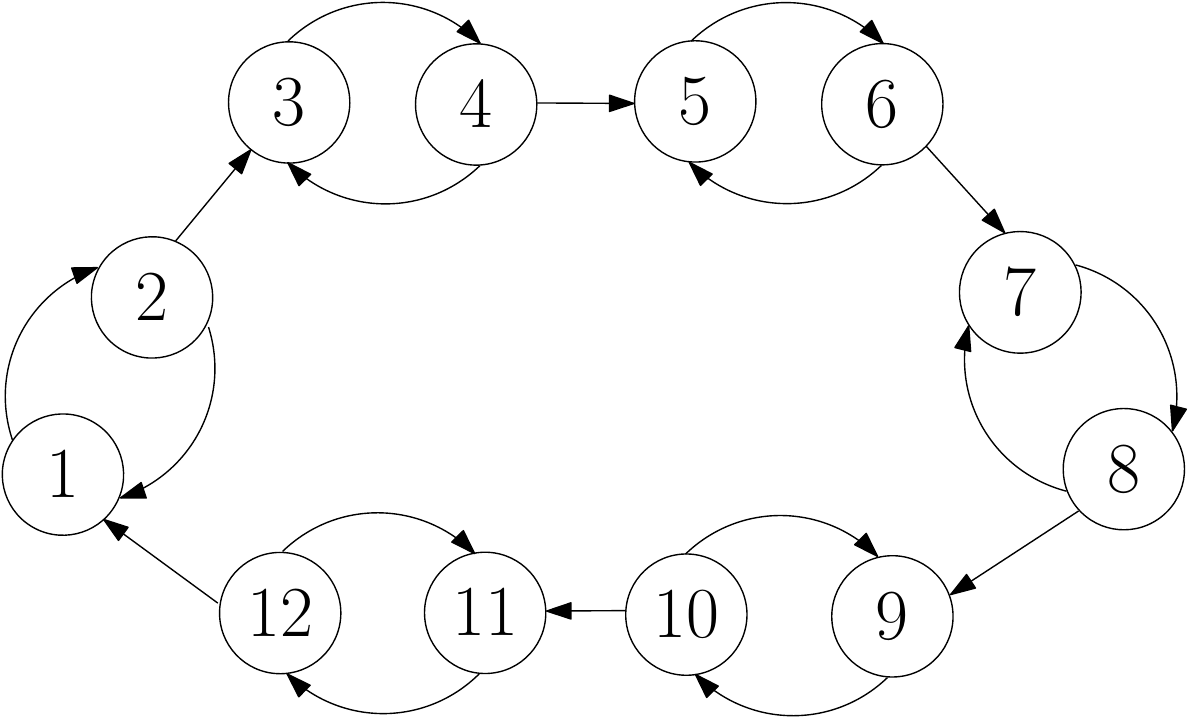}
\label{fig:fring2}
}
\end{minipage}
&
\begin{minipage}{2in}
\subtable[Results.]{
\begin{tabular}{|c|c|c|c|}
\hline
$N$ & $e_g$ & $D$ & Steps \\ \hline
2  &  1 &   1 &  18 \\ \hline
4  &  3 &   3 &  40 \\ \hline
6  &  5 &   5 &  62 \\ \hline
8  &  7 &   7 &  90 \\ \hline
10 &  9 &   9 & 122 \\ \hline
12 & 11 &  11 & 158 \\ \hline
14 & 13 &  13 & 198 \\ \hline
16 & 15 &  15 & 242 \\ \hline
18 & 17 &  17 & 290 \\ \hline
20 & 19 &  19 & 342 \\ \hline
\end{tabular}
\label{tbl:tring2}
}
\end{minipage}

\end{tabular}
\caption{Ring networks of size 2 rings.}
\label{fig:ring2}
\end{figure}
\begin{figure}[ht]
\centering
\begin{tabular}{cc}

\begin{minipage}{2in}
\subfigure[Sample network, $N=18$.]{
\includegraphics[scale=0.4]{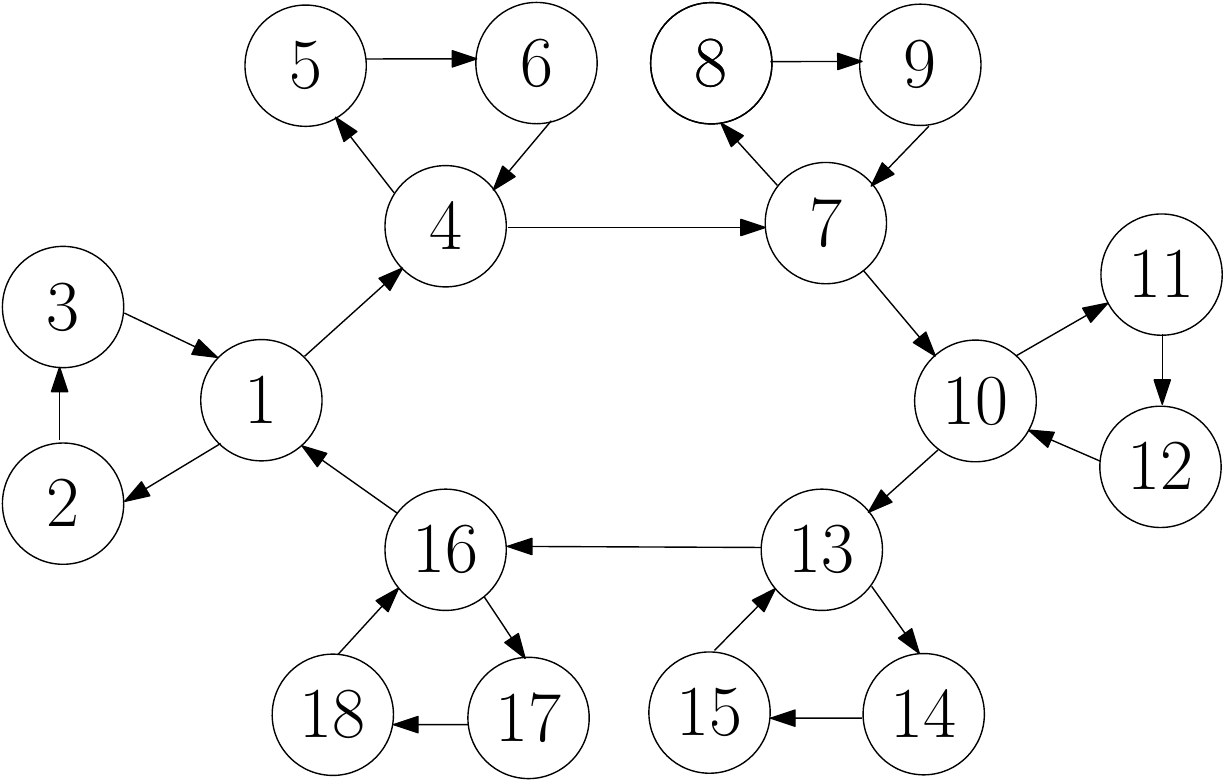}
\label{fig:fring3}
}
\end{minipage}
&
\begin{minipage}{2in}
\subtable[Results.]{
\begin{tabular}{|c|c|c|c|}
\hline
$N$ & $e_g$ & $D$ & Steps \\ \hline
5  & 4 & 4 &  57 \\ \hline
10 & 5 & 5 &  67 \\ \hline
15 & 6 & 6 &  78 \\ \hline
20 & 7 & 7 &  90 \\ \hline
25 & 8 & 8 & 101 \\ \hline
\end{tabular}
\label{tbl:tring3}
}
\end{minipage}

\end{tabular}
\caption{Ring networks of size 3 rings.}
\label{fig:ring3}
\end{figure}
\begin{figure}[ht]
\centering
\begin{tabular}{cc}

\begin{minipage}{2in}
\subfigure[Sample network, $N=14$.]{
\includegraphics[scale=0.4]{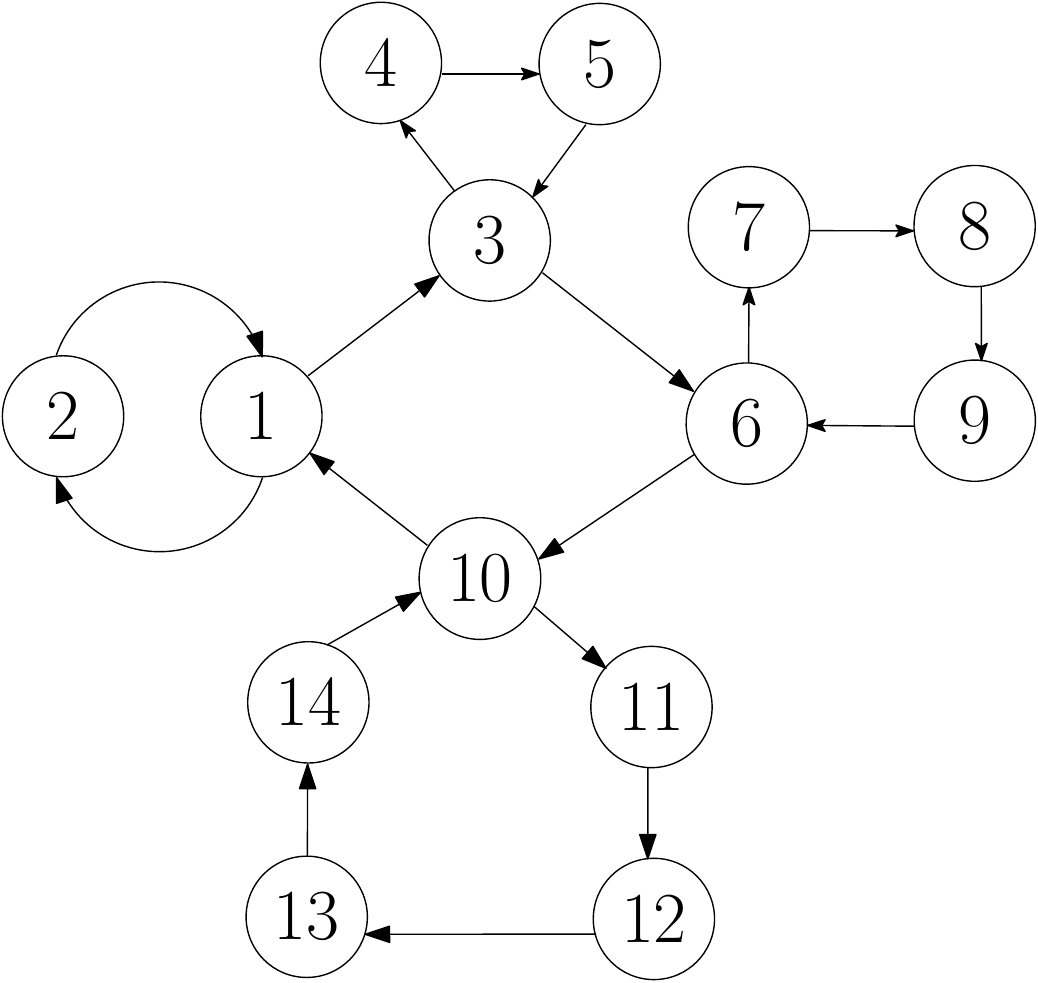}
\label{fig:fring4}
}
\end{minipage}
&
\begin{minipage}{2in}
\subtable[Results.]{
\begin{tabular}{|c|c|c|c|}
\hline
$N$ & $e_g$ & $D$ & Steps \\ \hline
2  & 1  & 1  &  18 \\ \hline
5  & 3  & 3  &  40 \\ \hline
9  & 5  & 5  &  63 \\ \hline
14 & 7  & 7  &  90 \\ \hline
20 & 9  & 9  & 121 \\ \hline
27 & 11 & 11 & 156 \\ \hline
\end{tabular}
\label{tbl:tring4}
}
\end{minipage}

\end{tabular}
\caption{Ring networks of increasing size rings.}
\label{fig:ring4}
\end{figure}

Besides our earlier example, we have empirically validated our solution in several test scenarios, with digraphs of different shapes and sizes, for example:

\begin{itemize}
\item A simple ring
(Figure~\ref{fig:ring1}).

\item A main ring linking a series of smaller rings of size two
(Figure~\ref{fig:ring2}).

\item A main ring linking a series of smaller rings of size three
(Figure~\ref{fig:ring3}).

\item A main ring linking a series of smaller rings of increasing size
(Figure~\ref{fig:ring4}).

\item A set of 11 random directed graphs, with up to 70 nodes each,
generated by the standard {\tt networkx} package.
\end{itemize}

The results support our claim for correctness and performance.


\section{Conclusions}
\label{sec:conclusion}

In this paper, we explicitly presented a first solution to the FSSP for
synchronous digraph-based P~systems with {\it simplex} channels.
Our design suggests, but does not need, ways to consolidate routing information
in such systems---this can be a topic for further study.

Our solution runs in $O(eD)$ steps and compares favourably,
i.e.~it is simpler and faster, than the best known solution for cellular automata~\cite{Ostrovsky95fastercomputation},
which runs in $O(ND)$ steps.
Taking into account the different problem constraints
and different computing capabilities of P~systems vs cellular automata,
we were expecting a simpler and a faster solution,
but not necessarily such a substantial speed improvement.
However, as noted before,
any performance comparison must be viewed with a grain of salt.

Our solution used a fixed size high-level rule set,
independent of the number of cells in the actual system
and of its structure.
This supports the case for reified cell IDs, complex symbols and generic rules
and suggests that such ingredients could be useful or even essential
in any distributed or just large system.


\section*{Acknowledgments}

The authors wish to acknowledge the contribution of Michael Dinneen, John Morris and Yun-Bum Kim, and the support of the University of Auckland FRDF grant 9843/3626216. The work of Florentin Ipate and Ionu\c{t} Mihai Niculescu was supported by CNCSIS-UEFISCSU, project number PNII-IDEI 643/2008.
The work of Cristian \c{S}tefan was supported by SOP-HRD grant 52826.



\nocite{DKN-UC2010,DKN-CMC-2011,DKN-JNC-2011}

\nocite{BernardiniGMV2008,AlhazovMV-WMC2008}

\nocite{Paun2000,Paun2010,MartinVidePPR2003,Paun2002}

\nocite{Moore1964,Goto1962,Waksman1966,Balzer1967,MooreL1968,SchmidW-IFIPTCS2004,Kobayashi1978,NishitaniH1981,Grefenstette1983,BerthiaumeBPSS2004,UmeoKNA2010}

\nocite{NishitaniH1981,Kobayashi1978,Even90computingwith,Ostrovsky95fastercomputation}


\bibliographystyle{abbrv}
\bibliography{MyRef}


\end{document}